\documentstyle[preprint,prabib,aps]{revtex}

\begin{document}
\draft
\date{\today}
\title{Warped Solitonic Deformations and Propagation
of Black Holes in 5D Vacuum Gravity }
\author{Sergiu I. Vacaru \thanks{
e--mail: sergiu$_{-}$vacaru@yahoo.com,\ vacaru@fisica.ist.utl.pt}}
\address{Physics Department, CSU Fresno,\ Fresno, CA 93740-8031, USA, \\
and \\
Centro Multidisciplinar de Astrofisica - CENTRA, Departamento de Fisica,\\
Instituto Superior Tecnico, Av. Rovisco Pais 1, Lisboa, 1049-001,\\
Portugal}
\author{D. Singleton \thanks{
e--mail: dougs@csufresno.edu}}
\address{Physics Department, CSU Fresno,\ Fresno, CA 93740-8031, USA}
\maketitle

\begin{abstract}
In this paper we use the anholonomic frames method
to construct exact solutions for vacuum 5D gravity with
metrics having off--diagonal components.
The solutions are in general anisotropic and possess
interesting features such as an anisotropic warp factor with respect to the
extra dimension, or a gravitational scaling/running of some of the physical
parameters associated with the solutions. A certain class of solutions are
found to describe Schwarzschild black holes which ``solitonically''
propagate in spacetime. The solitonic character of these black hole
solutions arises from the embedding of a 3D soliton configuration ({\it e.g.}
the soliton solutions to the Kadomtsev--Petviashvily or sine--Gordon
equations) into certain ansatz functions of the 5D metric. These
solitonic solutions may either violate or preserve local Lorentz
invariance. In addition there is a connection between these solutions
and noncommutative field theory.

Pacs 04.50.+h
\end{abstract}

\section{Introduction}

Recent work in string theory and brane physics has suggested new paradigms
for solving things like the hierarchy problem using novel treatments of the
extra dimensions that occur in these theories \cite{stringb}. One approach
is to view the extra dimension(s) as compactified, but at a size much larger
than the Planck length. This is the ``large extra dimension''
scenario \cite{arkani}. Another approach is to postulate that most particles
and fields are suppressed from moving into the extra dimension(s) by a
exponential ``warp'' factor with respect to the extra dimension(s). This is
the Randall-Sundrum (RS) model \cite{rs}.

The RS model in particular has gained considerable popularity in high energy
physics, cosmology and black hole physics
\cite{bh,csaki,csaki1,giannakis,kanti}.
Recently, it was shown \cite{vtheorem,v1} that by considering
off--diagonal metrics, which are diagonalized using anholonomic frames,
that the RS and Kaluza--Klein theories become locally anisotropic and
exhibit a variation or ``running'' of some of the constants associated with
these metrics. This variation of the effective 4D constants occurred with
respect to either the extra dimension coordinate or one of the angular
coordinates. These anisotropic solutions also exhibited RS style warp
factors, but without the need for any specific brane energy-momentum tensor
-- the warp factor arose from anisotropic, vacuum solutions in 5D gravity.

In refs. \cite{v} the anholonomic frames method was used to construct 4D and
5D vacuum solutions with one or more of the ansatz functions which
parameterized the off--diagonal metric being taken as a solitonic solution
to some nonlinear equation. The two cases considered were the solitonic
solutions of the 3D Kadomtsev--Petviashvili and sine--Gordon equations. (2D
solitonic solutions in 4D general relativity were originally introduced  in
ref. \cite{belinski}. A recent discussion can be found
in ref. \cite{belinski1}. Generalizations to
3D are given in refs. \cite{kad}). These 3D solitons,
embedded in the 5D spacetime, could be seen as solitonic, traveling pulses
which might have a larger cross section for being detected as compared to
standard gravitational waves.

In the present work we examine how black hole solutions are modified when
combined with the anisotropic/warped/solitonic configurations of refs. \cite
{v}. We begin by parameterizing a 5D spacetime with many off--diagonal
metric components. Into this off--diagonal metric we then embed a 4D
Schwarzschild solution. Adding to this various anisotropic/warped/solitonic
configurations it is found that the black hole solution can be modified in a
number of interesting ways. First, the horizon can be anisotropically
deformed. Along with the deformation of the horizon we find that some of the
``constants'' of the original Schwarzschild solution become dependent on one
or more of the coordinates ({\it e.g.} the mass of the solution becomes
dependent on the 5$^{th}$ coordinate. Second, the 3D solitonic
configurations can move or propagate the black hole either in the normal
uncompactified spatial dimensions, or in the extra, compactified spatial
dimension.

In this paper use the term ``locally anisotropic'' spacetime or
``anisotropic' spacetime'' for a 5D pseudo-Riemannian spacetime with an
anholonomic frame structure with mixed holonomic and anholonomic variables.
The anisotropy of the gravitational interactions arises as a result of the
off--diagonal metrics, or, equivalently, by their diagonalized versions
given with respect to anholonomic frames.

The paper has the following structure: in section II we present the
theoretical framework of the anholonomic frames method. We then introduce
the 5D metric ansatz form and write down the corresponding vacuum Einstein
equations. We show that it is possible to embed the Kadomtsev-Petviashvili
and sine-Gordon solitons into this system through one of the ansatz
functions. This creates a 3D gravitational soliton. In section III a 4D
Schwarzschild solution is embedded in the 5D spacetime and deformed in
various ways. These deformations lead to some of the ``constant'' parameters
of the original 4D Schwarzschild solution scaling with respect to some of
the coordinates. In section IV we focus on three classes of solutions with
small deformations of the horizon. The deformation of the horizons come from
the 3D soliton solutions of section II. In section V we examine solitonic
black hole solutions which move in the bulk 5D spacetime. These moving black
hole solutions can be given with either spherical or deformed horizons. In
section VI we discuss the physical features of the various solutions and
give our conclusions.

\section{Off--Diagonal Metrics and 3D solitons}

The Schwarzschild solution in {\it isotropic spherical coordinates}
\cite{ll} is given by
\begin{equation}
ds^{2}=\left( \frac{\widehat{\rho }-1}{\widehat{\rho }+1}\right)
^{2}dt^{2}-\rho _{g}^{2}\left( \frac{\widehat{\rho }+1}{\widehat{\rho }}
\right) ^4 \left( d\widehat{\rho }^{2}+\widehat{\rho }^{2}d\theta ^{2}+
\widehat{\rho }^{2}\sin ^{2}\theta d\varphi ^{2}\right) ,  \label{schw}
\end{equation}
The re--scaled isotropic radial coordinate is $\widehat{\rho } =\rho
/\rho _{g},$ with $\rho _{g}=r_{g}/4$; $\rho$ is connected with the usual
radial coordinate $r$ by $r=\rho \left( 1+r_{g}/4\rho \right) ^{2}$;
$r_{g}=2G_{[4]}m_{0}/c^{2}$ is the 4D Schwarzschild radius of a point
particle of mass $m_{0}$; $G_{[4]}=1/ M_{P[4]}^{2}$ is the 4D Newton
constant expressed via the Planck mass $M_{P[4]}$ ($M_{P[4]}$ may be an
effective 4D mass scale which arises from a more fundamental scale of the
full, higher dimensional spacetime). In the rest of the paper we set $c=1.$
The system of coordinates is $(t,\rho ,\theta ,\varphi ),$ where $t$ is the
time like coordinate and $(\rho ,\theta ,\varphi )$ are 3D spherical
coordinates. The metric (\ref{schw}) is a vacuum static solution of 4D
Einstein equations with spherical symmetry describing the gravitational
field of a point particle of mass $m_{0}.$ It has a singularity for $r=0$
and a spherical horizon at $r=r_{g},$ or at $\widehat{\rho }=1$ in the
re--scaled isotropic coordinates. This solution is parametrized by a
diagonal metric given with respect to holonomic coordinate frames. This
spherically symmetric solution can be deformed in various interesting ways
using the anholonomic frames method. Writing down a general off--diagonal
metric ansatz we first embed the above form of the Schwarzschild solutions
into the metric. Then this off--diagonal metric with the embedded
Schwarzschild solution is diagonalized with respect to the anholonomic
frames. The resulting field equations are of a form that is relatively
simple, and allows for analytical solutions which represent anisotropic
deformations of the original Schwarzschild solution.

\subsection{Off--diagonal metric ansatz}

We split the 5D coordinates $u^{\alpha }=(x^{i},y^{a})$ into  coordinates
$x^{i},$ with indices $i,j,k...=1,2,3,$ and coordinates $y^{a},$
with indices $a,b,c,...=4,5.$  Explicitly the coordinates are of
the form the coordinates
\[
x^{i}=(x^{1}=\chi, \qquad x^{2}=\lambda =\ln \widehat{\rho }, \qquad
x^{3}=\theta )\qquad \mbox{and }\qquad y^{a}=\left( y^{4}=v,
\qquad y^{5}=p\right) ,
\]
The metric interval is written as
\begin{equation}
ds^{2}=\Omega ^{2}(x^{i},v)\hat{{g}}_{\alpha \beta }\left( x^{i},v\right)
du^{\alpha }du^{\beta },  \label{cmetric}
\end{equation}
were the coefficients $\hat{{g}}_{\alpha \beta }$ are parametrized by the
ansatz {\scriptsize
\begin{equation}
\left[
\begin{array}{ccccc}
g_{1}+(w_{1}^{\ 2}+\zeta _{1}^{\ 2})h_{4}+n_{1}^{\ 2}h_{5} &
(w_{1}w_{2}+\zeta _{1}\zeta _{2})h_{4}+n_{1}n_{2}h_{5} & (w_{1}w_{3}+\zeta
_{1}\zeta _{3})h_{4}+n_{1}n_{3}h_{5} & (w_{1}+\zeta _{1})h_{4} & n_{1}h_{5}
\\
(w_{1}w_{2}+\zeta _{1}\zeta _{2})h_{4}+n_{1}n_{2}h_{5} & g_{2}+(w_{2}^{\
2}+\zeta _{2}^{\ 2})h_{4}+n_{2}^{\ 2}h_{5} & (w_{2}w_{3}++\zeta _{2}\zeta
_{3})h_{4}+n_{2}n_{3}h_{5} & (w_{2}+\zeta _{2})h_{4} & n_{2}h_{5} \\
(w_{1}w_{3}+\zeta _{1}\zeta _{3})h_{4}+n_{1}n_{3}h_{5} & (w_{2}w_{3}++\zeta
_{2}\zeta _{3})h_{4}+n_{2}n_{3}h_{5} & g_{3}+(w_{3}^{\ 2}+\zeta _{3}^{\
2})h_{4}+n_{3}^{\ 2}h_{5} & (w_{3}+\zeta _{3})h_{4} & n_{3}h_{5} \\
(w_{1}+\zeta _{1})h_{4} & (w_{2}+\zeta _{2})h_{4} & (w_{3}+\zeta _{3})h_{4}
& h_{4} & 0 \\
n_{1}h_{5} & n_{2}h_{5} & n_{3}h_{5} & 0 & h_{5}
\end{array}
\right] ,  \label{ansatzc}
\end{equation}
} The metric coefficients are smooth functions of the form:
\begin{eqnarray}
g_{1} &=&\pm 1,\qquad g_{2,3}=g_{2,3}(x^{2},x^{3}),\qquad
h_{4,5}=h_{4,5}(x^{i},v),  \nonumber \\
w_{i} &=&w_{i}(x^{i},v),\qquad n_{i}=n_{i}(x^{i},v),\qquad \zeta _{i}=\zeta
_{i}(x^{i},v),\qquad \Omega =\Omega (x^{i},v).  \nonumber
\end{eqnarray}
The quadratic line element (\ref{cmetric}) with metric coefficients
(\ref{ansatzc}) can be diagonalized,
\begin{equation}
\delta s^{2}=\Omega
^{2}(x^{i},v)[g_{1}(dx^{1})^{2}+g_{2}(dx^{2})^{2}+g_{3}(dx^{3})^{2}+h_{4}(
\hat{{\delta }}v)^{2}+h_{5}(\delta p)^{2}],  \label{cdmetric}
\end{equation}
with respect to the anholonomic co--frame $\left( dx^{i},\hat{{\delta }}
v,\delta p\right) ,$ where
\begin{equation}
\hat{\delta } v=dv+(w_{i}+\zeta _i)dx^i+\zeta _{5}\delta p\qquad
\mbox{ and } \qquad \delta p=dp+n_i dx^i  \label{ddif2}
\end{equation}
which is dual to the frame $\left( \hat{{\delta }}_{i},\partial _{4},\hat{{
\partial }}_{5}\right) ,$ where
\begin{equation}
\hat{{\delta }}_{i}=\partial _{i}-(w_{i}+\zeta _{i})\partial
_{4}+n_{i}\partial _{5},\qquad \hat{{\partial }}_{5}=\partial _{5}-\zeta
_{5}\partial _{4}.  \label{dder2}
\end{equation}
The simplest way to compute the nontrivial coefficients of the Ricci tensor
for the (\ref{cdmetric}) is to do this with respect to anholonomic bases (
\ref{ddif2}) and (\ref{dder2}), which reduces the 5D vacuum Einstein
equations to the following system:
\begin{eqnarray}
R_{2}^{2}=R_{3}^{3}=-\frac{1}{2g_{2}g_{3}}[g_{3}^{\bullet \bullet }-\frac{
g_{2}^{\bullet }g_{3}^{\bullet }}{2g_{2}}-\frac{(g_{3}^{\bullet })^{2}}{
2g_{3}}+g_{2}^{^{\prime \prime }}-\frac{g_{2}^{^{\prime }}g_{3}^{^{\prime }}
}{2g_{3}}-\frac{(g_{2}^{^{\prime }})^{2}}{2g_{2}}] &=&0,  \label{ricci1a} \\
R_{4}^{4}=R_{5}^{5}=-\frac{\beta }{2h_{4}h_{5}} &=&0,  \label{ricci2a} \\
R_{4i}=-w_{i}\frac{\beta }{2h_{5}}-\frac{\alpha _{i}}{2h_{5}} &=&0,
\label{ricci3a} \\
R_{5i}=-\frac{h_{5}}{2h_{4}}\left[ n_{i}^{\ast \ast }+\gamma n_{i}^{\ast }
\right]  &=&0,  \label{ricci4a}
\end{eqnarray}
with the conditions that
\begin{equation}
\Omega ^{q_{1}/q_{2}}=h_{4}~(q_{1}\mbox{ and }q_{2}\mbox{ are integers}),
\label{confq}
\end{equation}
and $\zeta _{i}$ satisfies the equations
\begin{equation}
\partial _{i}\Omega -(w_{i}+\zeta _{{i}})\Omega ^{\ast }=0,  \label{confeq}
\end{equation}
The coefficients of equations (\ref{ricci1a}) - (\ref{ricci4a}) are given by
\begin{equation}
\alpha _{i}=\partial _{i}{h_{5}^{\ast }}-h_{5}^{\ast }\partial _{i}\ln \sqrt{
|h_{4}h_{5}|},\qquad \beta =h_{5}^{\ast \ast }-h_{5}^{\ast }[\ln \sqrt{
|h_{4}h_{5}|}]^{\ast },\qquad \gamma =\frac{3h_{5}^{\ast }}{2h_{5}}-\frac{
h_{4}^{\ast }}{h_{4}}.  \label{abc}
\end{equation}
The various partial derivatives are denoted as ${\dot{a}}=\partial
a/\partial x^{1},a^{\bullet }=\partial a/\partial x^{2},a^{^{\prime
}}=\partial a/\partial x^{3},a^{\ast }=\partial a/\partial v.$ The details
of the computations leading to this system are given in refs. \cite
{vtheorem,v1}. This system of equations (\ref{ricci1a})--(\ref{ricci4a}),
(\ref{confq}) and (\ref{confeq}) can be solved by choosing one of the ansatz
functions ({\it e.g.} $h_{4}\left( x^{i},v\right) $ or $h_{5}\left(
x^{i},v\right) )$ to take some arbitrary, but physically interesting form.
Then the other ansatz functions can be analytically determined up to an
integration in terms of this choice. In this way one can generate many
solutions, but the requirement that the initial, arbitrary choice of the
ansatz functions be ``physically interesting'' means that one wants to make
this original choice so that the final solution generated in this way yield
a well behaved solution. To satisfy this requirement we start from well
known solutions of Einstein's equations and then use the above procedure to
deform this solutions in a number of ways.

The Schwarzschild solution is given in terms of the parameterization in
(\ref{ansatzc}) by
\begin{eqnarray}
g_{1} &=&\pm 1,\qquad g_{2}=g_{3}=-1,\qquad h_{4}=h_{4[0]}(x^{i}),\qquad
h_{5}=h_{5[0]}(x^{i}),  \nonumber \\
w_{i} &=&0,\qquad n_{i}=0,\qquad \zeta _{i}=0,\qquad \Omega =\Omega
_{\lbrack 0]}(x^{i}),  \nonumber
\end{eqnarray}
with
\begin{equation}
h_{4[0]}(x^{i})=\frac{b(\lambda )}{a(\lambda )},\qquad h_{5[0]}(x^{i})=-\sin
^{2}\theta ,\qquad \Omega _{\lbrack 0]}^{2}(x^{i})=a(\lambda )  \label{aux1}
\end{equation}
or alternatively, for another class of solutions,
\begin{equation}
h_{4[0]}(x^{i})=-\sin ^{2}\theta ,\qquad h_{5[0]}(x^{i})=\frac{b(\lambda )}{
a(\lambda )},  \label{aux2}
\end{equation}
were
\begin{equation}
a(\lambda )=\rho _{g}^{2}\frac{\left( e^{\lambda }+1\right) ^{4}}{
e^{2\lambda }}\qquad \mbox{ and }\qquad b(\lambda )=\frac{\left( e^{\lambda
}-1\right) ^{2}}{\left( e^{\lambda }+1\right) ^{2}},  \label{ab}
\end{equation}
Putting this together gives
\begin{equation}
ds^{2}=\pm d\chi ^{2}-a(\lambda )\left( d\lambda ^{2}+d\theta ^{2}+\sin
^{2}\theta d\varphi ^{2}\right) +b\left( \lambda \right) dt^{2}
\label{schw5}
\end{equation}
which represents a trivial embedding of the 4D Schwarzschild metric (\ref
{schw}) into the 5D spacetime. We now want to anisotropically deform the
coefficients of (\ref{schw5}) in the following way
\begin{eqnarray*}
h_{4[0]}(x^{i}) &\rightarrow &h_{4}(x^{i},v)=\eta _{4}\left( x^{i},v\right)
h_{4[0]}(x^{i}),\qquad h_{5[0]}(x^{i})\rightarrow h_{5}(x^{i},v)=\eta
_{5}\left( x^{i},v\right) h_{5[0]}(x^{i}), \\
\Omega _{\lbrack 0]}^{2}(x^{i}) &\rightarrow &\Omega ^{2}(x^{i},v)=\Omega
_{\lbrack 0]}^{2}(x^{i})\Omega _{\lbrack 1]}^{2}(x^{i},v),
\end{eqnarray*}
The factors $\eta _{i}$ and $\Omega _{\lbrack 1]}^{2}$ can be interpreted as
re-scaling or renormalizing the original ansatz functions. These
gravitational ``polarization'' factors -- $\eta _{4,5}$ and $\Omega
_{\lbrack 1]}^{2}$ -- generate non--trivial values for $
w_{i}(x^{i},v),n_{i}(x^{i},v)$ and $\zeta _{i}(x^{i},v),$ via the vacuum
equations (\ref{ricci1a})-- (\ref{confeq}). We shall also consider more
general nonlinear polarizations which can not be expresses as $h\sim $ $\eta
h_{[0]}.$ In the next section we show how the coefficients $a(\lambda )$ and
$b(\lambda )$ of the Schwarzschild metric can be polarized by choosing the
original, arbitrary ansatz function to be some 3D soliton configuration.

\subsection{3D soliton deformation of metric coefficients}

Vacuum gravitational 2D solitons in 4D Einstein vacuum gravity were
originally investigated by Belinski and Zakharov \cite{belinski}. In ref.
\cite{v} 3D solitonic configurations in 4D gravity were constructed on
anisotropic Taub-NUT backgrounds. Here we show that 3D solitonic
configurations can be embedded into 5D vacuum gravity.

The simplest way to construct a solitonic deformation of the off--diagonal
metric in equation (\ref{ansatzc}) is to take one of the ``polarization''
factors $\eta_{4}$, $\eta_{5}$ or the ansatz
function $n_{i}$ as a solitonic solution of some particular
non-linear equation. The rest of the ansatz functions can then be found by
carrying out the integrations of equations (\ref{ricci1a})-- (\ref{confeq}).

As an example of this procedure we take $\eta_{5}(r,\theta
,\chi )$ to be a soliton solution of the Kadomtsev--Petviashvili (KdP)
equation or (2+1) sine-Gordon (SG) equation (Refs. \cite{kad} contain the
original results, basic references and methods for handling such non-linear
equations with solitonic solutions). In the KdP case $\eta_{5}(v,\theta ,\chi )$
satisfies the following equation
\begin{equation}
\eta_{5}^{\ast \ast }+\epsilon \left( \dot{\eta}_{5}-6\eta_{5}\eta_{5}^{\prime
}+\eta_{5}^{\prime \prime \prime }\right) ^{\prime }=0,
\qquad \epsilon =\pm 1,
\label{kdp}
\end{equation}
while in the most general SG case $\eta_{5}(v,\chi )$ satisfies
\begin{equation}
\pm \eta_{5}^{\ast \ast } \mp \ddot{\eta}_{5}=
\sin (\eta_{5}).
\label{sineq}
\end{equation}
We can also consider less general versions of the SG equation where
$\eta _5$ depends on only one ({\it e.g.} $v$ and $x_1$) variable.
We will use the notation $\eta_{5}=\eta_{5}^{KP}$ or $\eta_{5}=\eta_{5}^{SG}$
($h_{5}=h_{5}^{KP}$ or $h_{5}=h_{5}^{SG}$) depending
on if ($\eta _5$ ) ($h_{5}$) satisfies equation (\ref{kdp}), or
(\ref{sineq}) respectively.

Having chosen a solitonic form for $h_{5}=h_{5}^{KP,SG},$ $h_{4}$ can be
found from
\begin{equation}
h_{4}=h_{4}^{KP,SG}=h_{[0]}^2 \left[
\left( {\sqrt{|h_{5}^{KP,SG} ( x^{i},v ) |}} \right) ^{\ast } \right] ^2
\label{p1b}
\end{equation}
where $h_{[0]}$ is a constant. By direct substitution it can be shown that
equation (\ref{p1b}) solves equation (\ref{ricci2a}) with $\beta $ given by (
\ref{abc}) when $h_{5}^{\ast }\neq 0$. If $h_{5}^{\ast }=0,$ then $h_{4}$ is
an arbitrary function $h_{4}(x^{i},v)$. In either case we will denote the
ansatz function determined in this way as $h_{4}^{KP,SG}$ although it does
not necessarily share the solitonic character of $h_{5}$. Substituting the
values $h_{4}^{KP,SG}$ and $h_{5}^{KP,SG}$ into $\gamma $ from equation (\ref
{abc}) gives, after two $v$ integrations of equation (\ref{ricci4a}), the
ansatz functions $n_{i}=n_{i}^{KP,SG}(v,\theta ,\chi )$. Here, for
simplicity, we set $g_{2,3}=-1$ so that the holonomic 2D background is
trivial. In ref. \cite{v} it was shown how to generate solutions using 2D
solitonic configurations for $g_{2}$ or $g_{3}$.

In addition to imposing a solitonic form on $h_{5}$ it is possible to carry
out a similar procedure starting with $h_{4}$ or $n_{i}$ ({\it i.e.} choose
$\eta_{4}$ or $n_{i}$ to satisfy equation (\ref{kdp}) or equation (\ref{sineq})
and then use equations (\ref{ricci1a})-- (\ref{confeq}), (\ref{abc}) and
(\ref{p1b}) and to determine the other ansatz functions).

The main conclusion of this subsection is that the ansatz (\ref{ansatzc}),
when treated with anholonomic frames, has a degree of freedom that allows
one to pick one of the ansatz functions ($\eta_{4}$ , $\eta_{5}$ , or $n_{i}$) to
satisfy some 3D solitonic equation. Then in terms of this choice all the
other ansatz functions can be generated up to carrying out some explicit
integrations and differentiations. In this way it is possible to build exact
solutions of the 5D vacuum Einstein equations with a solitonic character.

\section{Black Holes with Coordinate Dependent Parameters}

In this section exact 5D vacuum solutions are constructed by first embedding
the 4D spherically symmetric Schwarzschild solution into the 5D metric (\ref
{cmetric}). These solutions are then deformed by fixing $h_{4}$ or $h_{5}$
to be of the KdP or SG form discussed in the last subsection (in our example
we take $h_{5}$ to be of the solitonic form). This ``renormalizes'' the
gravitational constant, making it develop a coordinate dependence
\begin{equation}
\label{rhog}
\rho_{g}\rightarrow \overline{\rho }_{g}=
\omega \left( x^{i},v\right) \rho _{g}
\end{equation}
The polarization, $\omega \left( x^{i},v\right) $, is induced by the
solitonic character of $h_{4}$ or $h_{5}$ (in our example $h_{5}$). For
simplicity, we shall omit the indices $KP$ or $SG$ for the coefficients of
metric, since this will not result in ambiguities.

\subsection{Solutions with trivial conformal factor $\Omega =1$}

A particular case of metric (\ref{cdmetric}) with trivial conformal factor
is given by
\begin{eqnarray}
\delta s^{2} &=&[\pm d\chi ^{2}-d\lambda ^{2}-d\theta ^{2}+h_{4}(\hat{{
\delta }}v)^{2}+h_{5}(\delta p)^{2}],  \label{dm1} \\
\hat{{\delta }}v &=&dv+w_{i}\left( x^{i},v\right) dx^{i},\qquad \delta
p=dp+n_{i}\left( x^{i},v\right) dx^{i},  \nonumber
\end{eqnarray}
$h_{4}=\eta _{4}\left( x^{i},v\right) h_{4[0]}(x^{i})$ and $h_{5}=\eta
_{5}\left( x^{i},v\right) h_{5[0]}(x^{i})$. \ The functions $h_{4[0]}(x^{i})$
and $h_{5[0]}(x^{i})$ are taken as in (\ref{aux1}), or, inversely, as in
(\ref{aux2}). The polarizations, $\eta _{4,5}$ are chosen so $h_{5}$ is
a solution of either the KdP or SG soliton configuration, and that $h_{4}$
is then determined in terms of this choice. If $\omega ^{\ast }=\partial
_{v}\omega \neq 0,$ {\it i.e.} $h_{5}\neq 0,$ then the solution of equations
(\ref{ricci2a}) which are compatible with (\ref{ricci3a}) is
\begin{equation}
|h_{4}|=h_{[0]}^2 \left[ \left(
{\sqrt{|h_{5} ( x^{i},v ) |}} \right) ^{\ast } \right] ^2 ,
\label{p1}
\end{equation}
where $h_{[0]}=const$. Thus for $\eta _{5}^{\ast }\neq 0,$ the coefficients
$\eta _{4}$ and $\eta _{5}$ are related by
\[
| \eta _{4} ( x^{i},v ) h_{4(0)}(x^{i}) |
= h_{[0]}^{2}h_{5(0)}(x^{i})
\left[ \left( \sqrt{|\eta _{5} ( x^{i},v ) |} \right) ^{\ast }\right] ^2 .
\]
If $h_{5}^{\ast }=0$ ({\it i.e.} $\omega _{\varphi }^{\ast }=0$ and $\eta
_{5}^{\ast }=0$) then equation (\ref{ricci2a}) is solved by any $h_{4}$ as a
function of the variables $\left( x^{i},v\right) $ {\it i.e.} $h_{4}\left(
x^{i},v\right) $. For example, for $\eta _{5}=\omega ^{-2},$ $\omega ^{\ast
}\neq 0,$ and $h_{4,5(0)}$ from (\ref{aux1}) we can express the polarization
$\eta _{4}$ explicitly via functions $a,b$ and $\omega ,$
\begin{equation}
|\eta _{4}|=h_{[0]}^{2}\frac{b(\lambda )}{\sin ^{2}\theta a(\lambda )}\left[
\left( \omega ^{-1} ( x^{i},v ) \right) ^{\ast }\right] ^{2},  \label{aux3}
\end{equation}
which allows us to find \ the functional dependencies $h_{4,5}=h_{4,5}\left(
a,b,\omega \right) $ and to compute the coefficient $\gamma \left(
a,b,\omega \right) $ from (\ref{abc}).  After two integrations with respect
to $v$, the general solution of (\ref{ricci4a}), expressed via the
polarizations $\eta _{4}$ and $\eta _{5},$ is
\begin{eqnarray}
n_{k} &=&n_{k[1]} ( x^{i} ) +n_{k[2]} ( x^{i} ) \int
\frac{\eta _{4}}{(\sqrt{|\eta _{5}|})^{3}} dv,\qquad {\mbox {for}}
\;\;\eta _{5}^{\ast }\neq 0;  \label{nel} \\
&=&n_{k[1]} ( x^{i} ) +n_{k[2]} ( x^{i} ) \int \eta
_{4}dv,\qquad {\mbox {for}}\;\;\eta _{5}^{\ast }=0;  \nonumber \\
&=&n_{k[1]} ( x^{i} ) +n_{k[2]} ( x^{i} ) \int
\frac{1}{(\sqrt{|\eta _{5}|})^{3}}  dv,\qquad {\mbox {for}}\;\;\eta
_{4}^{\ast }=0,  \nonumber
\end{eqnarray}
where $n_{k[1,2]} ( x^{i} ) $ are fixed by boundary conditions. We
can simplify equations (\ref{nel}) by re-writing them in terms $\omega $.
For instance, if $\eta _{5}=\omega ^{-2},$ \ $\eta _{4}$ is taken from (\ref
{aux3}) and  $\omega ^{\ast }\neq 0,$ then one can write $n_{k}$ as
\[
n_{k}=n_{k[1]} ( x^{i} ) +n_{k[2]} ( x^{i} )
\int (\omega^* ( x^{i},v ))^2 \omega ^{-1} dv.
\]
The vacuum 5D metric (\ref{dm1}) constructed in this way is a particular
case of (\ref{ansatzc}) with $\Omega =1,w_{i}=0,$ $\zeta _{i}=0,$ but with
non--trivial values of $h_{4,5}=h_{4,5}\left( a,b,\omega \right) $ and
$n_{k}=n_{k}\left( x^{i},a,b,\omega \right) $. These non-trivial $h_{4,5}$
and $n_{k}$ give a renormalized gravitational constant of the form $\rho
_{g}\rightarrow \overline{\rho }_{g}=\omega \rho _{g}.$ In the trivial limit
$\omega \rightarrow 1,n_{k}\rightarrow 0,h_{4,5}\rightarrow h_{4,5(0)}(x^{i})
$ we recover not just the trivial embedding of the Schwarzschild metric into
the 5D spacetime (\ref{schw5}), but also the following metric
\begin{equation}
ds^{2}=\pm d\chi ^{2}-d\lambda ^{2}-d\theta ^{2}-\sin ^{2}\theta d\varphi
^{2}+\frac{b\left( \lambda \right) }{a(\lambda )}dt^{2}.  \label{schw5c}
\end{equation}
This metric is related to the metric in equation (\ref{schw5}) by
the multiplication of the 4D part of the metric by the conformal
factor $a^{-1}(\lambda )$ and a further extension to 5D. This
metric does not satisfy the vacuum 5D equations. Nevertheless, it
is possible to deform such a metric, by introducing
self--consistently some off--diagonal metric coefficients which
make the resulting metric into a solution of the vacuum Einstein
equations. The treatment of such off--diagonal metrics is that a
3D solitonic wave (satisfying the equation (\ref{kdp}), or
(\ref{sineq})) may polarize the constants of the conformally
transformed solution of equation (\ref{schw5c}) to generate a 5D
vacuum solution.

The metric in equation (\ref{dm1}) still has the same horizon and
singularities as the 4D Schwarzschild metric of equation (\ref{schw5}), but
it has the new feature that the gravitational constant is coordinate
dependent. This coordinate dependence comes from the coordinate dependence
of $\omega \left( x^{i},v\right) $.

There are two types of solutions which give an anisotropic renormalization
of the gravitational constant: one where the anholonomic coordinates are
chosen as $\left( v=\varphi ,p=t\right) $, and $\omega =\omega \left( \chi
,\lambda ,\varphi \right) $, $\eta _{5}=\omega ^{-2}\left( \chi
,\lambda ,\varphi \right) $, $h_{4[0]}=-\sin ^{2}\theta $ and
$h_{5[0]}=b/a$. These are called the $\varphi $-solutions. The second form is
with the anholonomic coordinates taken as $\left( v=t,p=\varphi \right) $,
and $\omega =\omega \left( \chi ,\lambda ,t\right) $, $\eta
_{5}=\omega ^{-2}\left( \chi ,\lambda ,t\right) $, $h_{4[0]}=b/a$
and $h_{5[0]}=-\sin ^{2}\theta )$. These are called the $t$-solutions. The
form of each of these solutions when $\omega ^{\ast }\neq 0$ is:
\begin{eqnarray}
\mbox{$\varphi$--solutions} &:&(x^{1}=\chi ,\qquad x^{2}=\lambda ,\qquad
x^{3}=\theta ,\qquad y^{4}=v=\varphi ,\qquad y^{5}=p=t),  \nonumber \\
g_{1} &=&\pm 1,\qquad g_{2}=g_{3}=-1,\qquad h_{4(0)}=-\sin ^{2}\theta
,\qquad h_{5(0)}=\frac{b(\lambda )}{a(\lambda )},  \nonumber \\
h_{4} &=&\eta _{4}(x^{i},\varphi )h_{4(0)}(x^{i}),\qquad h_{5}=\eta
_{5}(x^{i},\varphi )h_{5(0)}(x^{i}),\qquad \omega =\omega \left( \chi
,\lambda ,\varphi \right) ,  \nonumber \\
|\eta _{4}| &=&h_{(0)}^{2}\frac{b(\lambda )\omega ^{\ast 2}}{a(\lambda
)\omega ^{4}\sin ^{2}\theta },\qquad \eta _{5}=\omega ^{-2},  \nonumber \\
w_{i} &=&0,\qquad \zeta _{i}=0,\qquad n_{k}\left( x^{i},\varphi
\right) =n_{k[1]}(x^{i})+n_{k[2]}(x^{i})
\int (\omega^* ( x^{i}, \varphi ))^2 \omega ^{-1} d \varphi .  \label{sol5p1}
\end{eqnarray}
and
\begin{eqnarray}
\mbox{$t$--solutions} &:&(x^{1}=\chi ,\qquad x^{2}=\lambda ,\qquad
x^{3}=\theta ,\qquad y^{4}=v=t,\qquad y^{5}=p=\varphi ),  \nonumber \\
g_{1} &=&\pm 1,\qquad g_{2}=g_{3}=-1,\qquad h_{4(0)}=\frac{b(\lambda )}{%
a(\lambda )},\qquad h_{5(0)}=-\sin ^{2}\theta ,  \nonumber \\
h_{4} &=&\eta _{4}(x^{i},t)h_{4(0)}(x^{i}),\qquad h_{5}=\eta
_{5}(x^{i},t)h_{5(0)}(x^{i}),\qquad \omega =\omega \left( \chi
,\lambda ,t\right) ,  \nonumber \\
\eta _{4} &=&\omega ^{-2}\left( x^{i},t\right) ,\qquad \eta _{5}=\left[ \eta
_{5[0]}+ \frac{1}{h_{(0)}}\sqrt{\frac{b(\lambda )}{a(\lambda )}}
\int dt~\omega ^{-1}\left( x^{i},t\right) \right] ^{2},  \nonumber \\
w_{i} &=&0,\qquad \zeta _{i}=0,\qquad n_{k}=n_{k[1]}\left( x^{i}\right)
+n_{k[2]}\left( x^{i}\right) \int \eta _{4}|\eta _{5}|^{-3/2}dt.
\label{sol5t1}
\end{eqnarray}

As a specific example of the $\varphi$-solution we take $\eta _5$ to
satisfy the simplified SG equation -- $\ddot{\eta}_{5}=\sin (\eta_{5})
\rightarrow \partial _{\chi \chi} (\eta _5 ) = sin (\eta _5 )$.
The explicit form for $\eta _5$ is then
\begin{equation}
\eta _5 =4 \tan ^{-1} \left[ e ^{\pm\chi} \right]
\label{sol5p1a}
\end{equation}
In this case $\eta _5 ^* = \partial _{\varphi} \eta _5 =0$ so $h_4$ and
$\eta _4$ are arbitrary functions. We will take the simple case of
taking $\eta _4 =1$ so that $h_4$ retains its original form of
$-\sin ^2 \theta$. Also $n_k$ is now given by the middle form of
equation (\ref{nel}) and is thus fixed by the integration functions
$n_{k[1]} (x_i)$ and $n_{k[2]} (x_i)$. For simplicity we take these
functions to be zero so that $n_k =0$. Thus the metric of equation
(\ref{dm1}) takes the form
\begin{equation}
\label{sol5p1b}
\delta s^{2} =[\pm d\chi ^{2}-d\lambda ^{2}-d\theta ^{2}
-\sin ^2 \theta(\hat{{\delta }} v)^{2}+
4 \tan ^{-1} \left[ e ^{\pm\chi} \right]
\frac{b(\lambda )}{a(\lambda )}(\delta t)^{2}],
\end{equation}
An almost identical solution
can be constructed for the $t$-solutions, but with $\eta _4$ having
the form of the SG solution in equation (\ref{sol5p1a}).

If the internal coordinate $\chi$ is taken to be compactified as in
the standard Kaluza-Klein approach so that $\chi = 0$ is equivalent
to $\chi = 2 \pi$ then the SG form for $\eta _5$ or $\eta _4$ has the
problem of not being single valued ({\it i.e.} $\eta _5 (\chi =0) \ne
\eta _5 (\chi = 2 \pi )$). If one considers the extra dimension to be
uncompactified then one has the interesting feature that (via equation
(\ref{rhog}) -- $\overline{\rho }_{g}=\rho _{g}
(4 \tan ^{-1} \left[ e ^{\pm\chi} \right])^{-1/2}$)
the strength of the gravitational coupling decreases
as one moves off the 3D brane ($\chi =0$) into the extra dimension
($\chi \ne 0$) for the $+ \chi$ solution in equation (\ref{sol5p1a});
the gravitational coupling increases as one moves off the 3D brane
for the $-\chi$ solution. In order to explain why this extra, non-compactified
extra dimension is not observed one needs an overall RS-type \cite{rs}
exponential suppression as one moves in the extra dimension. We show
that this is possible in the next subsection.

\subsection{Solutions with non-trivial conformal factor $\Omega \neq 1$}

It is possible to generalize the ansatz of equation (\ref{dm1}) by taking a
nontrivial conformal factor $\Omega =\Omega _{\lbrack 0]} \Omega
_{\lbrack1]}.$ In this case the metric of (\ref{cdmetric}) takes the form
\begin{eqnarray}
\delta s^{2} &=&\Omega ^{2}[\pm d\chi ^{2}-d\lambda ^{2}-d\theta ^{2}+h_{4}
(\hat{{\delta }}v)^{2}+h_{5}(\delta p)^{2}],  \label{dm2} \\
\hat{{\delta }}v &=&dv+w_{i} ( x^{i},v ) dx^{i}+\zeta _{i} (
x^{i},v ) dx^{i},\qquad \delta p=dp+n_{i} ( x^{i},v ) dx^{i},
\nonumber
\end{eqnarray}
where $h_{4}=\eta _{4}(x^{i},t)h_{4(0)}(x^{i}),$ $h_{5}=\eta
_{5}(x^{i},t)h_{5(0)}(x^{i})$ and $\Omega _{\lbrack 0]}=\sqrt{a(\lambda )}
,\Omega _{\lbrack 1]}=\Omega _{\lbrack 1]} ( x^{k},v ) .$

The metric in (\ref{dm2}) satisfies the 5D vacuum Einstein equations if the
conformal factor $\Omega $ is related to the $h_{4}$ via equation (\ref
{confq}). This induces non-trivial values of $\zeta _{i}$ from
(\ref{confeq}). The factor $\Omega _{\lbrack 1]}$ is taken to be of the form
\[
\Omega _{\lbrack 1]} ( x^{k},v ) =\exp [-k|\chi |]\Omega _{\lbrack
2]} ( \lambda ,\theta ,v )
\]
{\it i.e.} it contains an exponential warp factor with respect to the 5$^{th}$
coordinate. Next we choose $h_{4}=\Omega ^{2}.$ This choice connects the
polarization $\eta _{4}$ with the conformal factor $\Omega _{\lbrack 2]},$
\begin{equation}
a(\lambda )\Omega _{\lbrack 2]}^{2}\left( \lambda ,\theta
,v\right) =\eta _{4}(x^{i},v)h_{4(0)}(x^{i}),  \label{pol1}
\end{equation}
The result is that $\zeta _{i}$ takes the form
\[
\zeta _{i}=\left( \Omega _{\lbrack 1]}^{\ast }\right) ^{-1}\partial
_{i}\Omega _{\lbrack 1]}+\left( \ln |\Omega _{\lbrack 1]}|\right) ^{\ast
}\partial _{i}\ln \sqrt{a(\lambda )}.
\]
The rest of the procedure of constructing $\varphi $--solutions and $t$
--solutions with a non--trivial conformal factor is similar to that from the
previous subsection. The form of these solutions with $\omega ^{\ast }\neq 0$
is
\begin{eqnarray}
\mbox{$\varphi$--solutions} &:&(x^{1}=\chi ,\qquad x^{2}=\lambda ,\qquad
x^{3}=\theta ,\qquad y^{4}=v=\varphi, \qquad y^{5}=p=t),  \nonumber \\
g_{1} &=&\pm 1,\qquad g_{2}=g_{3}=-1,\qquad h_{4(0)}=-\sin ^{2}\theta
,\qquad h_{5(0)}=\frac{b(\lambda )}{a(\lambda )},  \nonumber \\
h_{4} &=&\eta _{4}(x^{i},\varphi )h_{4(0)},\qquad h_{5}=\eta
_{5}(x^{i},\varphi )h_{5(0)},\qquad \omega =\omega \left( \chi
,\lambda ,\varphi \right) ,  \nonumber \\
\Omega  &=&\sqrt{a(\lambda )}\exp [-k|\chi |]\Omega _{\lbrack 2]}\left(
\lambda ,\theta ,\varphi \right) ,\qquad \Omega _{\lbrack 1]}=\exp [-k|\chi
|]\Omega _{\lbrack 2]}\left( \lambda ,\theta ,\varphi \right)
\label{sol5cp1} \\
\Omega _{\lbrack 2]}^{2} &=&\eta _{4}\left( x^{i},\varphi \right)
|h_{4(0)}|a^{-1}(\lambda ),  \nonumber \\
|\eta _{4}| &=&h_{(0)}^{2}\frac{b(\lambda ){\omega ^{\ast }}^{2}}{\omega
^{4}a(\lambda )\sin ^{2}\theta },\qquad \eta _{5}=\omega ^{-2},  \nonumber \\
w_{i} &=&0,\qquad n_{k}\left( x^{i},\varphi \right)
=n_{k[1]}(x^{i})+n_{k[2]}(x^{i})
\int (\omega^* ( x^{i}, \varphi ))^2 \omega ^{-1} d \varphi,  \nonumber \\
\zeta _{i} &=&\zeta _{i}(x^{i},\varphi )=\left( \Omega _{\lbrack 1]}^{\ast
}\right) ^{-1}\partial _{i}\Omega _{\lbrack 1]}+\left( \ln |\Omega _{\lbrack
1]}|\right) ^{\ast }\partial _{i}\ln \sqrt{a(\lambda )}  \nonumber
\end{eqnarray}
and
\begin{eqnarray}
\mbox{$t$--solutions} &:&(x^{1}=\chi ,\qquad x^{2}=\lambda ,\qquad
x^{3}=\theta ,\qquad y^{4}=v=t,\qquad y^{5}=p=\varphi ),  \nonumber \\
g_{1} &=&\pm 1,\qquad g_{2}=g_{3}=-1,\qquad h_{4(0)}=\frac{b(\lambda )}{%
a(\lambda )},\qquad h_{5(0)}=-\sin ^{2}\theta ,  \nonumber \\
h_{4} &=&\eta _{4}(x^{i},t)h_{4(0)},\qquad h_{5}=\eta
_{5}(x^{i},t)h_{5(0)},\qquad \omega =\omega \left( \chi ,\lambda
,t\right) ,  \nonumber \\
\Omega  &=&\sqrt{a(\lambda )}\exp [-k|\chi |]\Omega _{\lbrack 2]}\left(
\lambda ,\theta ,t\right) ,\qquad \Omega _{\lbrack 1]}=\exp [-k|\chi
|]\Omega _{\lbrack 2]}\left( \lambda ,\theta ,t\right)   \label{sol5ct1} \\
\Omega _{\lbrack 2]}^{2} &=&\eta _{4}\left( x^{i},t\right)
h_{4(0)}a^{-1}(\lambda ) ,  \nonumber \\
\eta _{4} &=&\omega ^{-2}(x^{i},t),\qquad \eta _{5}=\left[ \eta
_{5[0]}+ \frac{1}{h_{(0)}}\sqrt{\frac{b(\lambda )}{a(\lambda )}}
\int \omega ^{-1}(x^{i},t)dt \right] ^{2},  \nonumber \\
w_{i} &=&0,\qquad \zeta _{i}=0,\qquad n_{k}\left( x^{i},t\right)
=n_{k[1]}(x^{i})+n_{k[2]}(x^{i})\int \eta _{4}|\eta _{5}|^{-3/2}dt,
\nonumber \\
\zeta _{i} &=&\zeta _{i}(x^{i},t)=\left( \Omega _{\lbrack 1]}^{\ast }\right)
^{-1}\partial _{i}\Omega _{\lbrack 1]}+\left( \ln |\Omega _{\lbrack
1]}|\right) ^{\ast }\partial _{i}\ln \sqrt{a(\lambda )}.  \nonumber
\end{eqnarray}
It is straight forward to incorporate the special case solution
given in equation (\ref{sol5p1b}) of the last subsection to the
form in equation (\ref{sol5cp1}). The only change
is that the metric in equation (\ref{sol5p1b}) is multiplied by
an overall factor of $\exp[-2k|\chi |]$, which exponentially confines
all fields and particles (except gravity) to the 3D brane. Thus one
has the variation of the gravitational coupling discussed in the last
subsection along with a non-compact extra dimension. Because of the simply
form for $\eta _5$ for this special solution the $\zeta _i$ functions are
arbitrary functions rather than being given by the forms in equation
(\ref{sol5cp1}). The $t$-solution of the last
subsection can also be extended to the form given in
equation (\ref{sol5ct1}). Now the solitonic form for
$\eta_4$ ($\eta _4 =4 \tan ^{-1}[e^{\pm \chi}]$)
has an effect not only on the strength of the gravitational
coupling, but also on the overall conformal factor since
$\Omega ^2 \propto \eta_4$ from equation (\ref{sol5t1}). For the
$\eta _4 =4 \tan ^{-1}[e^{- \chi}]$ solution this just enhances
the exponential suppression as one moves away from $\chi =0$. For the
$\eta _4 =4 \tan ^{-1}[e^{+\chi}]$ solution the solitonic $\eta _4$
factor can initially dominate over the $e^{-2 k|\chi |}$ factor
as one first begins to move away from $\chi =0$.

The solutions of this section were generated by starting with a 4D
Schwarzschild metric and trivially embedding it into a 5D metric. It was
found that by fixing certain ansatz functions of the metric to take the form
of either the 3D KdP or SG soliton that the gravitational constant in 4D and
various other parameters of the metric became ``solitonically''
renormalized. In addition these 3D soliton configurations could propagate in
the 5D metric in such a manner so as to give a solitonic gravitational wave.

\section{Solitonic Deformations of Horizons}

By considering off--diagonal metrics and using the anholonomic frames method
it is possible to construct 4D and 5D black holes with non--spherical
horizons (In refs. \cite{vtheorem,v,v1} solutions with ellipsoidal,
toroidal, disk like, and bipolar horizons are discussed). In this section we
demonstrate that it is possible to deform spherical horizons by the
solitonic configurations of the last section.

We begin with the case of $t$--solutions which have a trivial polarization,
$\eta _{5}=1$, but with $\eta _{4}=\eta _{4}\left( x^{i},t\right)$, so that
$\eta _{4}^{\ast }\neq 0$. This will lead to solutions with a trivial
renormalization of constants, but with a small deformation of the spherical
horizon.

The spherical horizon of non--deformed solutions is defined by the condition
that the coefficient $b\left( \lambda \right) $ from equation (\ref{ab})
vanish. This occurs when $\lambda =0.$ A small deformation of the horizon
can be induced if $e^{\lambda }=e^{\epsilon \widetilde{\lambda }\left(
x^{i},t\right) }\simeq 1+\epsilon \widetilde{\lambda }\left( x^{i},t\right)
, $ for a small parameter $\epsilon \ll 1.$ This renormalizes $b\rightarrow
\widetilde{b}=b[1+\epsilon \eta \left( x^{i},t\right) ].$ Setting $\eta _{5}
= 1$ and $\eta _{4}=1+\epsilon \eta \left( x^{i},t\right)$ we require that
the pair $\left[\eta _{4}h_{4(0)},h_{5(0)}\right] $ satisfy equation (\ref
{ricci2a}). Since $h_5 = h_{5(0)} (x^i)$ so that $h_5^* =0$ this gives us
some freedom in choosing $\eta _{4}h_{4(0)}$. We use this freedom to take
$\eta \left( x^{i},t\right)$, as a solitonic solution of either the KdP
equation, (\ref{kdp}), or the SG equation ,(\ref{sineq}). This choice
induces a time dependent, solitonic deformation of the horizon. It is also
possible to generate solitonic deformations of the horizon which depends
on the extra spatial coordinate or an angular coordinate.
The ansatz functions
$n_{i}$ are determined from equation (\ref{nel}) with $\eta_{5}^{\ast }=0$
\[
n_{k}\left( x^{i},t\right) =n_{k[1]}\left( x^{i}\right) +n_{k[2]}\left(
x^{i}\right) \left[ t+\epsilon \int \eta \left( x^{i},t\right) dt\right] .
\]
The form for metrics of equation (\ref{dm1}) but with small deformations of
the horizon is
\begin{eqnarray}
\mbox{$t$--solutions} &:&(x^{1}=\chi , \qquad x^{2}=\lambda , \qquad
x^{3}=\theta, \qquad y^{4}=v=t, \qquad y^{5}=p=\varphi ),  \nonumber \\
g_{1}&=&\pm 1, \qquad g_{2} =g_{3}=-1, \qquad h_{4(0)}=\frac{b(\lambda )}{%
a(\lambda )}, \qquad h_{5(0)}= -\sin^2\theta ,  \nonumber \\
h_{4} &=&\eta _{4}(x^{i},t)h_{4(0)}(x^{i}), \qquad h_{5}=h_{5(0)}(x^{i}),
\qquad \eta _{4}=1+\epsilon \eta \left( x^{i},t\right) , \qquad \eta _{5}=
\omega =1,  \nonumber \\
w_{i} &=&0, \qquad \zeta _{i}=0, \qquad n_{k}\left( x^{i},t\right)
=n_{k[1]}(x^{i})+n_{k[2]}(x^{i})\left[ t+\epsilon \int \eta \left(
x^{i},t\right) dt\right] .  \label{sol5t2}
\end{eqnarray}
As a simple example of the above solution we can take $\eta $ as
satisfying the SG equation with only $t$ dependence: $\partial _{tt}
\eta  = \sin (\eta )$, which gives the solution
\begin{equation}
\label{sol5t2a}
\eta (t)= 4 \tan ^{-1} \left[e ^{\pm t}\right]
\end{equation}
If we take the additional simplifying assumption that
$n_{k[1]}(x^{i})=n_{k[2]}(x^{i})=0$ so that
$n_{k}( x^{i},t) =0$ then the metric from equation (\ref{dm1})
\begin{equation}
\delta s^{2} = [\pm d\chi ^{2}-d\lambda ^{2}-d\theta ^{2}+
(1 + \epsilon 4 \tan ^{-1} \left[e ^{\pm t}\right])
\frac{b (\lambda)}{a(\lambda )}
(\hat{{\delta }}t)^{2} - \sin ^2 \theta(\delta \varphi)^{2}],
\end{equation}
The $\eta _4 (t)= 1+ \epsilon 4 \tan ^{-1} \left[e ^{- t}\right]$
solution starts at $t=0$ with the finite value $1+ \epsilon \pi$ and
goes to $0$ as $t \rightarrow \infty$. Thus the horizon is initially
deformed, but approaches the Schwarzschild form as $t$ increases.
The $\eta _4 (t)= 1+ \epsilon 4 \tan ^{-1} \left[e ^{+ t}\right]$
solution also starts at $t=0$ with the value $1+ \epsilon \pi$,
but approaches the value $1 + 4 \epsilon \pi$. The deformation of
the horizon increases as $t$ increases.

It is possible to carry out a similar procedure on metrics of the
form (\ref{dm2}) which have a non--trivial conformal factor
\[
\Omega =\Omega _{\lbrack 0]}\exp [-k|\chi |]\Omega _{\lbrack 2]}\left(
\lambda ,\theta ,v\right) , \qquad h_{4}=\Omega ^{2}
\]
The conformal factor is connected with the polarization $\eta
_{4}=1+\epsilon \eta \left(x^{i},t\right)$, by equation (\ref{pol1}),
\[
a(\lambda )\exp [-2k|\chi |]\Omega _{\lbrack 2]}^{2}\left( \lambda ,\theta
,v\right) =\left[ 1+\epsilon \eta \left( x^{i},t\right) \right]
h_{4(0)}(x^{i}).
\]
The form of this solution is:
\begin{eqnarray}
\mbox{$t$--solutions} &:&(x^{1}=\chi , \qquad x^{2}=\lambda , \qquad
x^{3}=\theta, \qquad y^{4}=v=t,\qquad y^{5}=p=\varphi ),  \nonumber \\
g_{1}&=&\pm 1, \qquad g_{2} =g_{3}=-1, \qquad h_{4(0)}=\frac{b(\lambda )}{%
a(\lambda )}, \qquad h_{5(0)}= -\sin^2 \theta ,  \nonumber \\
h_{4} &=&\eta _{4}(x^{i},t)h_{4(0)}, \qquad h_{5}=h_{5(0)}, \qquad
\eta_{4}=1+\epsilon \eta \left( x^{i},t\right) , \qquad \eta _{5}= \omega =1,
\nonumber \\
\Omega &=&\sqrt{a(\lambda )}\exp [-k|\chi |]\Omega _{\lbrack 2]}\left(
\lambda ,\theta ,t\right) , \qquad \Omega _{\lbrack 1]}=\exp [-k|\chi
|]\Omega _{\lbrack 2]}\left( \lambda ,\theta ,t\right) ,  \nonumber \\
\Omega _{\lbrack 2]}^{2} &=&\eta _{4}\left( x^{i},t\right)
h_{4(0)}a^{-1}(\lambda ),  \label{sol5ct2} \\
w_{i} &=&0, \qquad \zeta _{i}=0, \qquad n_{k}\left( x^{i},t\right)
=n_{k[1]}(x^{i})+n_{k[2]}(x^{i})\left[ t+\epsilon \int \eta \left(
x^{i},t\right) dt\right] ,  \nonumber \\
\zeta _{i} &=&\zeta _{i}(x^{i},t)=\left( \Omega _{\lbrack 1]}^{\ast }\right)
^{-1}\partial _{i}\Omega _{\lbrack 1]}+\left( \ln |\Omega _{\lbrack
1]}|\right) ^{\ast }\partial _{i}\ln \sqrt{a(\lambda )}.  \nonumber
\end{eqnarray}
Since $h_{5}^{\ast }=0$ the function $\eta \left( x^{i},t\right)$ is
arbitrary and so many different possible scenarios exist for generating
small deformations of the horizon. Here we have taken $\eta \left(
x^{i},t\right)$ as being a 3D soliton configuration, leading to small,
solitonic deformations of the horizon. In appendix A we give the general
forms for these deformed horizon solutions with $\omega \ne 1$ and/or
$\Omega \ne 1$.

The solutions defined in this section and appendix A illustrate that extra
dimensional, anholonomic gravitational vacuum interactions can induce two
types of effects: renormalization of the physical constants and small
deformations of the horizons of black holes.

\section{Moving Soliton--Black Hole Configurations}

There is another class of vacuum solutions based on the 4D black hole
embedded into the 5D spacetime. It is possible to combine the solitonic
characteristics of the solutions discussed in section IIB with the embedded
4D black hole to generate a configuration which describes a 4D black hole
propagating in the 5D bulk. These solutions can be set up so as to have the
renormalization of the parameters of the metric and/or have deformed
horizons.

\subsection{The Schwarzschild black hole propagating as a 3D Soliton}

The horizon is defined by the vanishing of the coefficient $b\left( \lambda
\right)$ from equation (\ref{ab}). This occurs when $e^{\lambda }=1$. In
order to create a solitonically propagating black hole we define the
function $\tau =\lambda -\tau _{0}\left( \chi ,v\right)$, and let $%
\tau _{0}\left( \chi ,v\right) $ be a soliton solution of either the
3D KdP equation (\ref{kdp}), or the SG equation (\ref{sineq}). This
redefines $b\left( \lambda \right)$ as
\[
b\left( \lambda \right) \rightarrow B\left( x^{i},v\right) =\frac{e^{\tau }-1%
}{e^{\lambda }+1}.
\]
A class of 5D vacuum metrics of the form (\ref{dm1}) can be constructed by
parametrizing $h_{4}=\eta _{4}\left( x^{i},v\right) h_{4[0]}(x^{i})$ and $%
h_{5}=B\left(x^{i},v\right) /a\left( \lambda \right)$, or inversely, $%
h_{4}=B\left(x^{i},v\right) /a\left( \lambda \right)$ and $h_{5}=\eta
_{5}\left(x^{i},v\right) h_{5[0]}(x^{i}).$ The polarization $\eta
_{4}\left(x^{i},v\right) $ \ (or \ $\eta _{5} \left( x^{i},v\right) )$ is
determined from equation (\ref{p1})
\[
|\eta _{4}\left( x^{i},v\right) h_{4(0)}(x^{i})|=h_{[0]}^{2}\left[ \left(
\sqrt{\left| \frac{B\left( x^{i},v\right)}{a\left( \lambda \right)} \right|}%
\right) ^{\ast } \right] ^{2}
\]
or
\[
\left| \frac{B\left( x^{i},v\right)}{a\left( \lambda \right)} \right|
=h_{[0]}^{2}h_{5(0)}(x^{i}) \left[ \left( \sqrt{|\eta _{5}\left(
x^{i},v\right) |}\right) ^{\ast }\right] ^2.
\]
The last step in constructing of the form for these solitonically
propagating black hole solutions is to use $h_{4}$ and $h_{5}$ in equation (%
\ref{ricci4a}) to determine $n_k$
\begin{eqnarray}
n_{k} &=&n_{k[1]}( x^{i}) +n_{k[2]}( x^i ) \int
\frac{h_{4}}{(\sqrt{|h_{5}|})^3} dv, \qquad h_{5}^{\ast }\neq 0;
\label{nnn1} \\
&=&n_{k[1]}( x^{i}) +n_{k[2]}( x^{i}) \int
h_{4}dv,\qquad h_{5}^{\ast }=0;  \nonumber \\
&=&n_{k[1]}( x^{i}) +n_{k[2]}( x^{i}) \int
\frac{1}{(\sqrt{|h_{5}|})^3 } dv, \qquad h_{4}^{\ast }=0,  \nonumber
\end{eqnarray}
where $n_{k[1,2]}\left( x^{i}\right) $ are set by boundary conditions.

The simplest version of the above class of solutions are the $t$--solutions,
defined by a pair of ansatz functions, $\left[ B\left( x^{i},t\right)
,h_{5(0)}\right]$, with $h_{5}^{\ast }=0$ and $B\left( x^{i},t\right) $
being a 3D solitonic configuration. Such solutions have a spherical horizon
when $h_{4} =0$ {\it i.e.} when $\tau =0$. This solution describes
a propagating black hole horizon. The propagation occurs via a 3D
solitonic wave form depending on the time coordinate, $t$, and on
the 5$^{th}$ coordinate $\chi$. The form of the ansatz functions for this
solution (both with trivial and non-trivial conformal factors) is
\begin{eqnarray}
\mbox{$t$--solutions} &:&(x^{1}=\chi , \qquad x^{2}=\lambda , \qquad
x^{3}=\theta, \qquad y^{4}=v=t, \qquad y^{5}=p=\varphi ),  \nonumber \\
g_{1}&=&\pm 1, \qquad g_{2}=g_{3}=-1, \qquad \tau = \lambda -\tau _{0}\left(
\chi ,t\right), \qquad h_{5(0)}=-\sin ^{2}\theta ,  \nonumber \\
h_{4} &=&B/a(\lambda ), \qquad h_{5}=h_{5(0)}(x^{i}), \qquad \omega = \eta
_{5}=1, \qquad B\left( x^{i},t\right) =\frac{e^{\tau }-1}{e^{\lambda }+1},
\nonumber \\
w_{i} &=&\zeta _{i}=0, \qquad n_{k}\left( x^{i},t\right)
=n_{k[1]}(x^{i})+n_{k[2]}(x^{i})\int B\left( x^{i},t\right) dt  \label{sol6t}
\end{eqnarray}
and
\begin{eqnarray}
\mbox{$t$--solutions} &:&(x^{1}=\chi , \qquad x^{2}=\lambda , \qquad
x^{3}=\theta, \qquad y^{4}=v=t, \qquad y^{5}=p=\varphi ),  \nonumber \\
g_{1}&=&\pm 1, \qquad g_{2}=g_{3}=-1, \qquad \tau = \lambda -\tau _{0}\left(
\chi ,t\right), \qquad h_{5}=h_{5(0)}=-\sin ^{2}\theta ,  \nonumber
\\
h_{4} &=&\frac{B}{a(\lambda )},\qquad B\left( x^{i},t\right) = \frac{e^{\tau
}-1}{e^{\lambda}+1}, \qquad \Omega _{\lbrack 2]}^{2}=Ba^{-1}(\lambda )\exp
[2k|\chi |],  \nonumber \\
\Omega &=&\sqrt{a(\lambda )}\exp [-k|\chi |]\Omega _{\lbrack 2]}\left(
\lambda ,\theta ,t\right) , \qquad \Omega _{\lbrack 1]}=\exp [-k|\chi
|]\Omega_{\lbrack 2]}\left( \lambda , \theta ,t\right) ,  \nonumber \\
w_{i} &=&\zeta _{i}=0, \qquad n_{k}\left( x^{i},t\right)
=n_{k[1]}(x^{i})+n_{k[2]}(x^{i})\int B\left( x^{i},t\right) dt,  \nonumber \\
\zeta _{i} &=&\zeta _{i}(x^{i},t)=\left( \Omega _{\lbrack 1]}^{\ast }\right)
^{-1}\partial _{i}\Omega _{\lbrack 1]}+\left( \ln |\Omega _{\lbrack
1]}|\right) ^{\ast }\partial _{i}\ln \sqrt{a(\lambda )}.  \label{sol6ct}
\end{eqnarray}
These forms are similar to the solutions in (\ref{sol5t2}) and
(\ref{sol5ct2})).

As a simple example of the above solutions we take $\tau _0$ to
satisfy the SG equation $\partial _{\chi \chi} \tau _0
- \partial_{t t} \tau _0 = \sin (\tau _0 )$. This has the
standard propagating kink solution
\begin{equation}
\label{sol6ta}
\tau _0 (\chi , t) = 4 \tan ^{-1}
\left[ \pm \gamma (\chi - V t) \right]
\end{equation}
where $\gamma = (1 - V^2)^{-1/2}$ and $V$ is
the velocity at which the kink moves into the extra dimension
$\chi$. To obtain the simplest form of this solution we also
take $n_{k[1]}(x^{i})=n_{k[2]}(x^{i}) = 0$ as in our previous
examples. This example can be easily extended to the solution
in equation (\ref{sol6ct}) with a non-trivial conformal factor
that gives an exponentially suppressing factor, $\exp[-2 k
| \chi | ]$. In this manner one has an effective 4D black hole
which propagates from the 3D brane into the non-compact, but
exponentially suppressed extra dimension, $\chi$.

In appendix B we give the forms for two classes of solutions
with $\Omega =1$ and for two classes with $\Omega \neq 1$. These solutions
contain various combinations of the previous solutions: solitonically
propagating horizons; deformed horizons; renormalized constants.

The solutions constructed in this section and appendix B describe propagating 4D
Schwarzschild black holes in a bulk 5D spacetime. The propagation arises
from requiring that certain of the ansatz functions take a 3D soliton form.
In the simplest version of these propagating solutions the parameters of the
ansatz functions are constant, and the horizons are spherical. It was also
shown that such propagating solutions could be formed with a renormalization
of the parameters and/or deformation of the horizons.

\section{Conclusions and Discussion}

In this work, we have addressed the issue of constructing various classes of
exact solutions which generalize the 4D Schwarzschild black hole metric to a
bulk 5D vacuum spacetime.

Throughout the paper the method of anholonomic frames was applied to a
general, off--diagonal 5D metric ansatz in order to integrate the Einstein
equations and determine the ansatz functions. In this way a host of
different solutions were investigated which had a number of interesting
features such as a ``renormalization'' of the parameters of the ansatz,
deformation of the horizons, and self--consistent solitonic propagation of
black holes in the bulk 5D spacetime. The essential features of each of
these solutions would remain unchanged even if we had started with
ellipsoidal or toroidal rather than spherical horizons (see ref. \cite
{vtheorem} for discussions on black holes with ellipsoidal or toroidal
horizons), introduced matter sources, considered cosmological constants, or
used an anisotropic Taub NUT background \cite{v,v1}.

In the first part of the paper, we considered solutions with renormalized
``constants''-- $\rho _{g}\rightarrow \overline{\rho }_{g}=\omega \rho _{g}$
where $\rho _{g}=r_{g}/4$ and $r_{g}$ is the Schwarzschild radius. The
induced coordinate dependence of $\rho _g$ was connected with the extra
dimension and/or the anholonomic gravitational interaction. Since $\rho _g$
depends on Newton's constant, the value of the point mass, and the speed of
light one could interpret this induced coordinate dependence as a variation
of Newton's constant, an anisotropic mass, or a variation in the speed of
light/gravity waves. In the first case we obtain a Randall-Sundrum like
correction of Newton's force law \cite{vtheorem,v1}. In our case this arises
from the use of anholonomic frames instead of a brane configurations. In the
second case, we obtain an anisotropic gravitational interaction coming from
the anisotropic effective mass. In the third case, we can have metrics which
violate local Lorentz symmetry since light and gravitational waves could
have propagation speeds which are not constant \cite{csaki}.

In the second part of the paper, we analyzed configurations where 3D
solitons in 5D vacuum gravity induced small ``solitonic'' deformations of
the horizon. The simplest examples were ``pure'' deformations which did not
renormalize the constants. It was possible to combine the horizon
deformation effects with the renormalization of the constants considered in
the first part of the paper. The anisotropy of these configurations was
connected with either an angular coordinate or the time-like coordinate.

The third part of the paper was devoted to solutions which described 4D
black holes moving solitonically in the 5D spacetime. The solitonic
character of these solutions came from setting the form of one of the metric
ansatz functions to either the 3D KdP or SG soliton configuration. It was
shown that these propagating black hole solutions could be combined with the
effects discussed in the first to parts of the paper, generating propagating
black holes with renormalized constants and/or deformed horizons.

Mathematically this paper demonstrates the usefulness of the anholonomic
frames method, developed in refs. \cite{vtheorem,v,v1}, in constructing
solutions to the 4D and 5D Einstein vacuum field equations with
off--diagonal metrics. By introducing anholonomic transforms we were able to
diagonalize the metrics and simplify the system of equations in terms of the
ansatz functions. These solutions describe a generic anholonomic
(anisotropic) dynamics coming from the off--diagonal metrics. They also
generalize the class of exact 4D solutions with linear extensions to the
bulk 5D gravity given in ref. \cite{giannakis}.

We emphasize that the constructed solutions do not violate the conditions of
the Israel and Carter theorems \cite{israel} on spherical symmetry of black
hole solutions in asymptotically flat spacetimes which were formulated and
proved for 4D spacetimes.

Salam, Strathee and Peracci \cite{sal} explored the association between
gauge fields and the coefficients of off--diagonal metrics in extra
dimensional gravity. In the present paper we have used anholonomic frames
with associated nonlinear connections of 5D and 4D (pseudo) Riemannian
spaces \cite{vtheorem,v,v1}, to find anisotropic solutions with running
constants. These solutions may point to Lorentz violations effects in
Yang--Mills and electrodynamic theories induced from 5D vacuum gravity.
(Refs. \cite{kost} gives a general analysis of the problem for Planck--scale
physics and string theory). In addition the conditions of frame anholonomy
can be associated with a specific noncommutative relation. This indicates
that these solutions with generic anisotropy may mimic violations of Lorentz
symmetry similar to that found in noncommutative field theories \cite{ncom}.

Our results should be compared with the recent work of ref. \cite{kanti}
where the possibility of obtaining localized black hole solutions in brane
worlds was investigated. In ref. \cite{kanti} this was accomplished by
introducing a dependence of the 4D line element on extra dimension. It was
concluded that for either an empty bulk or a bulk containing scalar or gauge
fields that no conventional type of matter could support such a dependence.
For a particular diagonal ansatz for the 5D line--element it was found that
an exotic, shell--like, distribution of matter was required. The
off--diagonal metrics and anholonomic frames considered in the present work
are more general and get around the restrictions in \cite{kanti}. We find
that it is possible to generate warped factors and/or anisotropies in 5D
vacuum gravity by taking off--diagonal metrics and using the anholonomic
frames method. It is not necessary to generate these effects by some brane
configurations with specific energy--momentum tensors.

As a concluding remark, we note that the black hole solutions of the type
considered in this work suggest that the localizations to 4D in extra
dimension gravity depends strongly not only on the nature of the bulk matter
distribution, but also on the type of 5D metrics used: for diagonal metrics
one needs exotic bulk matter distributions in order to localize black hole
solutions; for off--diagonal metrics with associated anholonomic frames, one
can construct localized black hole solutions with various effect
(renormalization of constants, deformed horizons, solitonically propagating
black hole configurations) by considering an anholonomic 5D vacuum
gravitational dynamics. Matter distribution, and gravitational and matter
field interactions on an effective 4D spacetime can be modeled as induced by
off--diagonal metrics and anholonomic frame dynamics in a vacuum extra
dimension gravity.

\subsection*{Acknowledgments}

S.\ V. thanks C. Grojean and I. Giannakis for useful observations and
discussions on anisotropic extra dimension black hole solutions. The work is
supported by a 2000--2001 California State University Legislative Award and
a NATO/Portugal fellowship grant at the Technical University of Lisbon.

\appendix

\section{Solitonic deformations of horizons with non-trivial
polarization and conformal factor}

In this appendix we collect the expressions for the solitonically
deformed horizons with non-trivial polarizations and/or conformal
factors.

\subsection{$\omega \neq 1$ and $\Omega =1$}

The form of solutions (\ref{sol5p1}) and (\ref{sol5t1}) can be modified so
as to allow configurations with both renormalizations of constants and small
deformations of the horizon. The simplest way of doing this is to introduce
horizon deformations into $\eta _{4}$ and $\eta _{5}$ via the modification
of $\omega $ and $b$ as
\begin{equation}
\widetilde{\omega }^{-2}=\omega ^{-2}\left[ 1+\epsilon \eta \left(
x^{i},v\right) \right] \mbox{ and }\widetilde{b}=b\left[ 1+\epsilon \eta
\left( x^{i},v\right) \right] .  \label{ren5}
\end{equation}
The remaining calculations leading to the final form of the ansatz functions
are similar to those in subsections IIB and IIIA. We will just write down
the final form of the ansatz functions
\begin{eqnarray}
\mbox{$\varphi$--solutions} &:&(x^{1}=\chi ,\qquad x^{2}=\lambda ,\qquad
x^{3}=\theta ,\qquad y^{4}=v=\varphi ,\qquad y^{5}=p=t),  \nonumber \\
g_{1} &=&\pm 1,\qquad g_{2}=g_{3}=-1,\qquad h_{4(0)}=-\sin ^{2}\theta
,\qquad h_{5(0)}=\frac{b(\lambda )}{a(\lambda )},  \nonumber \\
h_{4} &=&\eta _{4}(x^{i},\varphi )h_{4(0)}(x^{i}),\qquad h_{5}=\eta
_{5}(x^{i},\varphi )h_{5(0)}(x^{i}),  \label{sol5p3} \\
|\eta _{4}| &=&h_{(0)}^{2}\frac{\widetilde{b}(\widetilde{\omega }^{\ast
})^{2}}{\widetilde{\omega }^{4}a(\lambda )\sin ^{2}\theta },\qquad \eta _{5}=%
\widetilde{\omega }^{-2},\qquad \omega =\omega \left( \chi ,\lambda
,\varphi \right) ,  \nonumber \\
\widetilde{\omega }^{-2} &=&\omega ^{-2}\left[ 1+\epsilon \eta \left(
x^{i},\varphi \right) \right] ,\qquad \widetilde{b}=b\left[ 1+\epsilon \eta
\left( x^{i},\varphi \right) \right] ,  \nonumber \\
w_{i} &=&0,\qquad \zeta _{i}=0,\qquad n_{k}\left( x^{i},\varphi \right)
=n_{k[1]}(x^{i})+n_{k[2]}(x^{i})
\int (\widetilde{\omega}^* ( x^{i}, \varphi ))^2 \omega ^{-1}
d \varphi . \nonumber
\end{eqnarray}
and
\begin{eqnarray}
\mbox{$t$--solutions} &:&(x^{1}=\chi ,\qquad x^{2}=\lambda ,\qquad
x^{3}=\theta ,\qquad y^{4}=v=t,\qquad y^{5}=p=\varphi ),  \nonumber \\
g_{1} &=&\pm 1,\qquad g_{2}=g_{3}=-1,\qquad h_{4(0)}=\frac{b(\lambda )}{%
a(\lambda )},\qquad h_{5(0)}=-\sin ^{2}\theta ,  \nonumber \\
h_{4} &=&\eta _{4}(x^{i},t)h_{4(0)}(x^{i}),\qquad h_{5}=\eta
_{5}(x^{i},t)h_{5(0)}(x^{i}),\qquad \omega =\omega \left( \chi
,\lambda ,t\right) ,  \nonumber \\
\eta _{4} &=&\widetilde{\omega }^{-2}\left( x^{i},t\right) ,\qquad \eta _{5}=%
\left[ \eta _{5[0]}+\frac{1}{h_{(0)}}\sqrt{\frac{\widetilde{b}}{a}}\int dt~%
\widetilde{\omega }^{-1}\left( x^{i},t\right) \right] ^{2},  \nonumber \\
\widetilde{\omega }^{-2} &=&\omega ^{-2}\left[ 1+\epsilon \eta \left(
x^{i},t\right) \right] ,\qquad \widetilde{b}=b\left[ 1+\epsilon \eta \left(
x^{i},t\right) \right] ,  \label{sol5t3} \\
w_{i} &=&0,\qquad \zeta _{i}=0,\qquad n_{k}\left( x^{i},t\right)
=n_{k[1]}(x^{i})+n_{k[2]}(x^{i})\int \eta _{4}|\eta _{5}|^{-3/2}dt.
\nonumber
\end{eqnarray}
In the limit $\epsilon \rightarrow 0,$ $\eta _{4}=\eta _{5}=1$ and $n_{k}=0$
and we obtain a 4D metric which is a conformally transformed version of the
Schwarzschild metric in equation (\ref{schw5c}).

\subsection{$\omega \neq 1$ and $\Omega \neq 1$}

In a similar manner we can construct metrics with deformed horizons and a
non-trivial conformal factor. Taking effective polarizations as in (\ref
{ren5}) and recalculating the ansatz functions in equations (\ref{sol5cp1})
and (\ref{sol5ct1}) generates metrics of the form (\ref{dm2}) with deformed
horizons. With $\omega ^{\ast }\neq 0$ the form for these solutions is
\begin{eqnarray}
\mbox{$\varphi$--solutions} &:&(x^{1}=\chi ,\qquad x^{2}=\lambda ,\qquad
x^{3}=\theta ,\qquad y^{4}=v=\varphi ,\qquad y^{5}=p=t),  \nonumber \\
g_{1} &=&\pm 1,\qquad g_{2}=g_{3}=-1,\qquad h_{4(0)}=-\sin ^{2}\theta
,\qquad h_{5(0)}=\frac{b(\lambda )}{a(\lambda )},  \nonumber \\
h_{4} &=&\eta _{4}(x^{i},\varphi )h_{4(0)},\qquad h_{5}=\eta
_{5}(x^{i},\varphi )h_{5(0)},\qquad \omega =\omega \left( \chi
,\lambda ,\varphi \right) ,  \nonumber \\
\Omega  &=&\sqrt{a(\lambda )}\exp [-k|\chi |]\Omega _{\lbrack 2]}\left(
\lambda ,\theta ,\varphi \right) ,\qquad \Omega _{\lbrack 1]}=\exp [-k|\chi
|]\Omega _{\lbrack 2]}\left( \lambda ,\theta ,\varphi \right) ,  \nonumber \\
\Omega _{\lbrack 2]}^{2} &=&\eta _{4}\left( x^{i},\varphi \right)
|h_{4(0)}|a^{-1}(\lambda ),\qquad \widetilde{b}=b\left[
1+\epsilon \eta \left( x^{i},\varphi \right) \right] ,  \nonumber \\
|\eta _{4}| &=&h_{(0)}^{2}\frac{\widetilde{b}{\widetilde{\omega }^{\ast 2}}}{%
\widetilde{\omega }^{2}a(\lambda )\sin ^{2}\theta },\qquad \eta _{5}=%
\widetilde{\omega }^{-2},\qquad \widetilde{\omega }^{-2}=\omega ^{-2}\left[
1+\epsilon \eta \left( x^{i},\varphi \right) \right] ,  \nonumber \\
w_{i} &=&0,\qquad n_{k}\left( x^{i},\varphi \right)
=n_{k[1]}(x^{i})+n_{k[2]}(x^{i})
\int (\widetilde{\omega}^* ( x^{i}, \varphi ))^2 \omega ^{-1} d \varphi ,
\label{sol5p4} \\
\zeta _{i} &=&\zeta _{i}(x^{i},\varphi )=\left( \Omega _{\lbrack 1]}^{\ast
}\right) ^{-1}\partial _{i}\Omega _{\lbrack 1]}+\left( \ln |\Omega _{\lbrack
1]}|\right) ^{\ast }\partial _{i}\ln \sqrt{a(\lambda )}  \nonumber
\end{eqnarray}
and
\begin{eqnarray}
\mbox{$t$--solutions} &:&(x^{1}=\chi ,\qquad x^{2}=\lambda ,\qquad
x^{3}=\theta ,\qquad y^{4}=v=t,\qquad y^{5}=p=\varphi ),  \nonumber \\
g_{1} &=&\pm 1,\qquad g_{2}=g_{3}=-1,\qquad h_{4(0)}=\frac{b(\lambda )}{%
a(\lambda )},\qquad h_{5(0)}=-\sin ^{2}\theta ,  \nonumber \\
h_{4} &=&\eta _{4}(x^{i},t)h_{4(0)},\qquad h_{5}=\eta
_{5}(x^{i},t)h_{5(0)},\qquad \omega =\omega \left( \chi ,\lambda
,t\right) ,  \label{sol5t4} \\
\Omega  &=&\sqrt{a(\lambda )}\exp [-k|\chi |]\Omega _{\lbrack 2]}\left(
\lambda ,\theta ,t\right) ,\qquad \Omega _{\lbrack 1]}=\exp [-k|\chi
|]\Omega _{\lbrack 2]}\left( \lambda ,\theta ,t\right)   \nonumber \\
\Omega _{\lbrack 2]}^{2} &=&\eta _{4}\left( x^{i},t\right)
|h_{4(0)}|a^{-1}(\lambda ),\qquad \widetilde{\omega }%
^{-2}=\omega ^{-2}\left[ 1+\epsilon \eta \left( x^{i},t\right) \right] ,
\nonumber \\
\eta _{4} &=&\widetilde{\omega }^{-2}(x^{i},t),\qquad \eta _{5}=\left[ \eta
_{5[0]}+ \frac{\sqrt{\widetilde{b}}}{h_{(0)}\sqrt{a(\lambda )}}\int
\widetilde{\omega }^{-1}(x^{i},t)dt \right] ^{2}, \qquad
w_{i} =\zeta _{i}=0, \nonumber \\
n_{k}\left( x^{i},t\right)
&=&n_{k[1]}(x^{i})+n_{k[2]}(x^{i})\int \eta _{4}|\eta _{5}|^{-3/2}dt, \qquad
\widetilde{b}=b\left[ 1+\epsilon \eta \left( x^{i},t\right) \right] ,
\nonumber \\
\zeta _{i} &=&\zeta _{i}(x^{i},t)=\left( \Omega _{\lbrack 1]}^{\ast }\right)
^{-1}\partial _{i}\Omega _{\lbrack 1]}+\left( \ln |\Omega _{\lbrack
1]}|\right) ^{\ast }\partial _{i}\ln \sqrt{a(\lambda )}.  \nonumber
\end{eqnarray}
The solutions constructed in this subsection, (\ref{sol5p4}) and (\ref
{sol5t4}), reduce to the trivial embedding of the Schwarzschild metric in
the 5D spacetime (\ref{schw}) in the limit when $\epsilon \rightarrow 0,$
$\eta _{4}=\eta _{5}=1$ and $n_{k}=0$.

\section{Solitonic propagating black holes with deformed horizons and/or
non-trivial conformal factor}

In this appendix we collect the expressions for the solitonically
propagating black holes with deformed horizons and/or non-trivial
conformal factors.

\subsection{Solitonic propagating and deformed black holes, $\Omega =1$}

To the solitonically moving black holes of section V we add the
effects of renormalization of constants and horizon deformation found in
sections III and IV. This is accomplished by modifying the function $B\left(
x^{i},v\right) $ and $\omega$ in a manner similar to (\ref{ren5})
\[
\widetilde{\omega }^{-2}=\omega ^{-2}\left[ 1+\epsilon \eta \left(
x^{i},v\right) \right] \qquad \mbox{ and } \qquad \widetilde{B}=B\left(
x^{i},v\right) \left[ 1+\epsilon \eta \left( x^{i},v\right) \right]
\]
and, for $\varphi $--solutions, write
\[
h_{4}=\eta _{4}\left( x^{i},\varphi \right) h_{4(0)} \qquad \mbox{ and }
\qquad h_{5}= \frac{\widetilde{B}\left( x^{i},\varphi \right)} {%
a\left(\lambda \right)\widetilde{\omega }^2} ,
\]
or for $t$--solutions
\[
h_{4}=\frac{\widetilde{B}\left( x^{i},t\right)} {a\left(\lambda \right)
\widetilde{\omega }^2} \qquad \mbox{ and } \qquad h_{5}=\eta _{5}\left(
x^{i},t\right) h_{5(0)},
\]
The polarization $\eta _{4}\left( x^{i},\varphi \right) , $ or $\eta
_{5}\left( x^{i},t\right) ,$ is found by equation (\ref{p1}), and the
coefficients $n_{i4}\left( x^{i},v\right) $ are found by integrating as in
equation (\ref{nnn1}) for $h_{4}^{\ast }\neq 0$ and $h_{5}^{\ast}\neq 0.$
The forms for exact solutions are
\begin{eqnarray}
\mbox{$\varphi$--solutions} &:&(x^{1}=\chi , \qquad x^{2}=\lambda , \qquad
x^{3}=\theta, \qquad y^{4}=v=\varphi , \qquad y^{5}=p=t),  \nonumber \\
g_{1}&=&\pm 1, \qquad g_{2}=g_{3}=-1, \qquad h_{4(0)}=-\sin ^{2}\theta ,
\qquad h_{4}=\eta_{4}(x^{i},\varphi )h_{4(0)}(x^{i}),  \nonumber \\
h_{5} &=&\frac{\widetilde{B}\left( x^{i},\varphi \right) }
{a\left( \lambda \right)\widetilde{\omega }^2} ,
\qquad |\eta _{4}|=\frac{h_{(0)}^2}{a(\lambda )\sin ^2 \theta}
\left[ \left( \frac{\sqrt{\widetilde{B}}}{\widetilde{\omega }}
\right)^{\ast} \right]^2,  \label{sol7p} \\
\omega &=&\omega \left( \chi ,\lambda ,\varphi \right) , \qquad
\tau=\lambda -\tau _{0}\left( \chi ,\varphi \right) , \qquad
\widetilde{\omega }^{-2}=\omega ^{-2}\left[ 1+\epsilon \eta \left(
x^{i},\varphi \right) \right] ,  \nonumber \\
\widetilde{B} &=&B\left[ 1+\epsilon \eta \left( x^{i},\varphi \right) \right]
,\qquad B\left( x^{i},\varphi \right) =\frac{\left( e^{\tau }-1\right)}{
\left(e^{\lambda }+1\right)} ,  \nonumber \\
w_{i} &=&\zeta _{i}=0, \qquad n_{k}\left( x^{i},\varphi \right)
=n_{k[1]}(x^{i})+n_{k[2]}(x^{i})\int \frac{h_{4}}{(\sqrt{|h_{5}|})^{3}}
d\varphi .
\nonumber
\end{eqnarray}
and
\begin{eqnarray}
\mbox{$t$--solutions} &:&(x^{1}=\chi , \qquad x^{2}=\lambda , \qquad
x^{3}=\theta,\qquad y^{4}=v=t, \qquad y^{5}=p=\varphi ),  \nonumber \\
g_{1}&=&\pm 1, \qquad g_{2}=g_{3}=-1, \qquad h_{5(0)}=-\sin ^{2}\theta ,
\qquad h_{5}=\eta_{5}(x^{i},t)h_{5(0)}(x^{i}),  \nonumber \\
h_{4} &=&\frac{\widetilde{B}\left( x^{i},t\right) }
{a\left(\lambda \right)\widetilde{\omega }^2} ,
\qquad \eta _{5}=\left[ \eta _{5[0]}+
\frac{1}{h_{(0)} \sqrt{a(\lambda )}} \int dt \frac{\sqrt{\widetilde{B}}}{
\widetilde{\omega } \left( x^i,t\right) } \right] ^2,  \nonumber \\
\omega &=&\omega \left( \chi ,\lambda ,t\right) , \qquad \tau
=\lambda -\tau_{0}\left( \chi ,t\right) , \qquad \widetilde{\omega }
^{-2}= \omega ^{-2}\left[ 1+\epsilon \eta \left( x^{i},t\right) \right] ,
\nonumber \\
\widetilde{B} &=&B\left[ 1+\epsilon \eta \left( x^{i},t\right) \right]
,\qquad B\left( x^{i},t\right) = \frac{\left( e^{\tau }-1 \right) }{\left(
e^{\lambda}+1\right) },  \label{sol7c} \\
w_{i} &=&\zeta _{i}=0, \qquad n_{k}\left( x^{i},t\right)
=n_{k[1]}(x^{i})+n_{k[2]}(x^{i})\int \frac{h_4}{(\sqrt{|h_{5}|})^3} dt.
\nonumber
\end{eqnarray}
In the limits $\epsilon \rightarrow 0,$ $\eta _{4}=\eta _{5}=1,B\rightarrow
b $ and $n_k=0$ we obtain a 4D metric trivially extended into 5D, which is a
conformally transformed version of the Schwarzschild metric (\ref{schw5c}).

\subsection{Solitonic propagating and deformed black holes, $\Omega \neq 1$}

We can revise (\ref{sol7p}) and (\ref{sol7c}) so as to generate metrics of
the class (\ref{dm2}), but with nontrivial conformal factors. The forms for
these solutions with $\omega ^{\ast }\neq 0$ are
\begin{eqnarray}
\mbox{$\varphi$--solutions} &:&(x^{1}=\chi ,\qquad x^{2}=\lambda , \qquad
x^{3}=\theta, \qquad y^{4}=v=\varphi , \qquad y^{5}=p=t),  \nonumber \\
g_{1}&=&\pm 1, \qquad g_{2}=g_{3}=-1, \qquad h_{4(0)}=-\sin ^{2}\theta ,
\qquad h_{4}=\eta_{4}(x^{i},\varphi ) h_{4(0)}(x^{i}),  \nonumber \\
h_{5} &=&\frac{\widetilde{B}\left( x^{i},\varphi \right)} {\widetilde{\omega
}^2 a\left( \lambda \right)} , \qquad |\eta _{4}|=\frac{h_{(0)}^2}{a(\lambda
)\sin ^{2}\theta} \left[ \left( \frac{\sqrt{\widetilde{B}}}{\widetilde{%
\omega }} \right)^{\ast } \right] ^2 ,  \nonumber \\
\omega &=&\omega \left( \chi ,\lambda ,\varphi \right) , \qquad
\tau=\lambda -\tau _{0}\left( \chi ,\varphi \right) , \qquad
\widetilde{\omega }^{-2}=\omega ^{-2} \left[ 1+\epsilon \eta \left(
x^{i},\varphi \right) \right] ,  \nonumber \\
\widetilde{B} &=&B\left[ 1+\epsilon \eta \left( x^i,\varphi \right) \right]
, \qquad B\left( x^{i},\varphi \right) = \frac{\left( e^{\tau }-1\right)}{%
\left( e^{\lambda}+1\right)} ,  \nonumber \\
\Omega &=&\sqrt{a(\lambda )}\exp [-k|\chi |]\Omega _{\lbrack 2]}\left(
\lambda ,\theta ,\varphi \right) ,\qquad \Omega _{\lbrack 1]}=\exp
[-k|\chi|]\Omega _{\lbrack 2]} \left( \lambda ,\theta ,\varphi \right) ,
\nonumber \\
\Omega _{\lbrack 2]}^{2} &=&\eta _{4}\left( x^{i},\varphi \right)
|h_{4(0)}|a^{-1}(\lambda ),  \label{sol8p} \\
w_{i} &=&\zeta _{i}=0, \qquad n_{k} ( x^{i},\varphi )
=n_{k[1]}(x^{i})+n_{k[2]}(x^{i})\int \frac{h_{4}}{(\sqrt{|h_{5}|})^3}
d\varphi ,  \nonumber \\
\zeta _{i} &=&\zeta _{i}(x^{i},\varphi )=\left( \Omega _{\lbrack 1]}^{\ast
}\right) ^{-1}\partial _{i}\Omega _{\lbrack 1]}+\left( \ln |\Omega _{\lbrack
1]}|\right) ^{\ast }\partial _{i}\ln \sqrt{a(\lambda )}  \nonumber
\end{eqnarray}
and
\begin{eqnarray}
\mbox{$t$--solutions} &:&(x^{1}=\chi , \qquad x^{2}=\lambda , \qquad
x^{3}=\theta, \qquad y^{4}=v=t, \qquad y^{5}=p=\varphi ),  \nonumber \\
g_{1}&=&\pm 1, \qquad g_{2}=g_{3}=-1, \qquad h_{5(0)}=-\sin ^{2}\theta ,
\qquad h_{5}=\eta_{5}(x^{i},t)h_{5(0)}(x^{i}),  \nonumber \\
h_{4} &=&\frac{\widetilde{B}\left( x^i,t\right) }{\widetilde{\omega }^2
a\left(\lambda \right)} , \qquad \eta _{5}=\left[ \eta _{5[0]}+
\frac{1}{h_{(0)} \sqrt{a(\lambda )}} \int dt \frac{\sqrt{\widetilde{B}}}{%
\widetilde{\omega } \left( x^i,t\right)} \right] ^2,  \nonumber \\
\omega &=&\omega \left( \chi ,\lambda ,t\right) ,\qquad \tau
=\lambda -\tau_{0}\left( \chi ,t\right) , \qquad \widetilde{\omega }%
^{-2}=\omega ^{-2} \left[ 1+\epsilon \eta \left( x^{i},t\right) \right] ,
\nonumber \\
\widetilde{B} &=&B\left[ 1+\epsilon \eta \left( x^{i},t\right) \right] ,
\qquad B\left( x^{i},t\right) =\frac{\left( e^{\tau }-1\right)} {\left(
e^{\lambda}+1\right)} ,  \nonumber \\
\Omega &=&\sqrt{a(\lambda )}\exp [-k|\chi |]\Omega _{\lbrack 2]}\left(
\lambda ,\theta ,t\right) , \qquad \Omega _{\lbrack 1]}=\exp [-k|\chi
|]\Omega _{\lbrack 2]}\left( \lambda ,\theta ,t\right) ,  \nonumber \\
\Omega _{\lbrack 2]}^{2} &=&\eta _{4}\left( x^{i},t\right)
|h_{4(0)}|a^{-1}(\lambda ),  \label{sol8c} \\
w_{i} &=&\zeta _{i}=0, \qquad n_{k}\left( x^{i},t\right)
=n_{k[1]}(x^{i})+n_{k[2]}(x^{i})\int \frac{h_4}{(\sqrt{|h_{5}|})^3 }
dt.  \nonumber
\end{eqnarray}
The solutions of equations (\ref{sol8p}) and (\ref{sol8c}) reduce to the
trivial embedding of the 4D Schwarzschild metric into 5D spacetime from
equation (\ref{schw}) in the limit $\epsilon \rightarrow 0,$ $\eta _{4}=\eta
_{5}=1,B\rightarrow b$ and $n_k=0$

\end{document}